\documentclass[
prb,
reprint,
superscriptaddress,
showpacs,
amsmath,amssymb,
floatfix,
]{revtex4-1}
\usepackage{dcolumn}% Align table columns on decimal point
\usepackage{bm}% bold math
\usepackage{turnstile}

\usepackage{graphicx}
\usepackage{amsmath,amssymb}
\usepackage{url}
\usepackage{multirow}
\usepackage{float}
\usepackage{xcolor}
\DeclareMathOperator{\re}{\mathop{\mathrm{Re}}}

\newcommand{\Eq}[1]{Eq.~(\ref{#1})}
\newcommand{\Eqs}[1]{Eqs.~(\ref{#1})}
\begin{document}

\title{ Anomalous current-voltage characteristics of SFIFS Josephson junctions with weak ferromagnetic interlayers
}
\author{T.~Karabassov}
\email{iminovichtair@gmail.com}
\affiliation{National Research University Higher School of Economics, 101000 Moscow, Russia}

\author{A.~V.~Guravova}
\affiliation{National Research University Higher School of Economics, 101000 Moscow, Russia}

\author{A.~Yu.~Kuzin}
\affiliation{Skolkovo Institute of Science and Technology, 121205 Moscow, Russia}
\affiliation{Department of Physics, Moscow State Pedagogical University, 119992 Moscow, Russia}

\author{E.~A.~Kazakova}
\affiliation{National Research University Higher School of Economics, 101000 Moscow, Russia}

\author{S.~Kawabata}
\affiliation{National Institute of Advanced Industrial Science and Technology,
1-1-1 Umezono, Tsukuba, Ibaraki 305-8563, Japan}

\author{B.~G.~Lvov}
\affiliation{National Research University Higher School of Economics, 101000 Moscow, Russia}

\author{A.~S.~Vasenko}
\affiliation{National Research University Higher School of Economics, 101000 Moscow, Russia}
\affiliation{I.E. Tamm Department of Theoretical Physics, P.N. Lebedev Physical Institute, Russian Academy of Sciences, 119991 Moscow, Russia}

\begin{abstract}
We present a quantitative study of the current-voltage characteristics (CVC) of SFIFS Josephson junctions (S denotes bulk superconductor, F - metallic ferromagnet, I - insulating barrier) with weak ferromagnetic interlayers in the diffusive limit. The problem is solved in the framework of the nonlinear Usadel equations. We consider the case of a strong tunnel barrier such that the left SF and the right FS bilayers are decoupled. We calculate the density of states (DOS) in SF bilayers using a self-consistent numerical method. Then we obtain the CVC of corresponding SFIFS junctions, and discuss their properties for different set of parameters including the thicknesses of ferromagnetic layers, the exchange field, and the magnetic scattering time. We observe the anomalous nonmonotonic CVC behavior in case of weak ferromagnetic interlayers, which we ascribe by DOS energy dependencies in case of small exchange fields in F layers.
\end{abstract}

\maketitle

%%%%%%%%%%%%%%%%%%%%%%%%%%%%%%%%%%%%%%%%%%%%%%%%%%%%%%%%%%%%%%%%%%%%%%%%%%%%
%%%%%%%%%%%%%%%%%%%%%%%%%%%%%%%%%%%%%%%%%%%%%%%%%%%%%%%%%%%%%%%%%%%%%%%%%%%%

\section{Introduction}

%%%%%%%%%%%%%%%%%%%%%%%%%%%%%%%%%%%%%%%%%%%%%%%%%%%%%%%%%%%%%%%%%%%%%%%%%%%%
%%%%%%%%%%%%%%%%%%%%%%%%%%%%%%%%%%%%%%%%%%%%%%%%%%%%%%%%%%%%%%%%%%%%%%%%%%%%

It is well known that superconductivity and ferromagnetism are two competing antagonistic  orders. In superconductors (S) electrons form Cooper pairs with opposite spins and momenta, while in ferromagnetic metals (F) electron spins tend to align in parallel. Nevertheless, it is possible to combine in one hybrid structure the S and F layers, which leads to observation of many striking phenomena. The reason is the superconducting proximity effect, i.e. the superconducting correlations leakage into a ferromagnetic metal due to the Andreev reflection processes. \cite{BuzdinRMP, GolubovRMP, BergeretRMP, Demler1997, Ozaeta2012R, Bergeret2013, Bobkova2017} As a consequence, the real part of the pair wave function performs the damped oscillatory behavior in a ferromagnetic metal. Hence, since the oscillations are spatially dependent, it is possible to realize a transition from ``0'' to ``$\pi$'' phase state in S/F/S structures upon changing the F layer thickness.\cite{BuzdinRMP} The proximity effect is characterized by two length scales of decay and oscillations of the real part of the pair wave function in a ferromagnetic layer, $\xi_{f1}$ and $\xi_{f2}$, correspondingly.\cite{BuzdinRMP} If we consider the exchange field $h$ as the only important parameter of a ferromagnetic material, both lengths are equal to $\xi_h = \sqrt{D_f/h}$, where $D_f$ is the diffusion constant in the ferromagnetic metal.

The existence of such phenomena makes possible the creation of so-called Josephson $\pi$ junctions with a negative critical current.\cite{BuzdinRMP, GolubovRMP} Oscillations of the pair wave function in the F layer leads to several interesting phenomena in S/F/(S) systems, including nonmonotonic critical temperature dependence,\cite{Jiang1995, Izyumov2002, Fominov2002, Khaydukov2018, Karabassov2019} Josephson critical current oscillations,\cite{Buzdin1982, Buzdin1991, Ryazanov2001, Blum2002, Sellier2004, Bauer2004, Bell2005, Oboznov2006, Shelukhin2006, Vasenko2008, Anwar2010, Khaire2010, Robinson2010, Baker2014, Halterman2014, Loria2015, Bakurskiy2017, Yamashita2017, Kontos2002, Guichard2003, Born2006, Pepe2006, Weides2006, Weides2006_2, Pfeiffer2008, Bannykh2009, Kemmler2010} and density of states (DOS) oscillations.\cite{Buzdin2000, Kontos2001, Halterman2004, Vasenko2011} S/F hybrid structures have many promising applications in single flux quantum (SFQ) circuits,\cite{Hilgenkamp2008, Shafranjuk2016} spintronic devices,\cite{Linder2015} like memory elements\cite{Larkin2012, Golovchanskiy2016, Bakurskiy2016, Soloviev2017, Caruso2018, Golovchanskiy2018_1, Bakurskiy2018, Nevirkovets2018, Nevirkovets2018_2, Shafraniuk2019} and spin-valves,\cite{Tagirov1999, Alidoust2015, Halterman2016_1, Halterman2016_2, Srivastava2017, Halterman2018, Alidoust2018} magnetoelectronics,\cite{Baek2014, Gingrich2016, Golovchanskiy2018_2} qubits,\cite{Feofanov2010} artificial neural networks,\cite{Soloviev2018} microrefrigerators,\cite{Ozaeta2012, Kawabata2013} low-temperature sensitive electron thermometers,\cite{Giazotto2015} etc.

However, junctions with a ferromagnetic interlayer as well as other normal metal junctions (for example, SFNFS), proposed as elements of novel superconducting nanoelectronics, have limited applicability since such junctions have low resistance values.\cite{Bell2004, book}
This situation is resolved by addition of an insulating barrier (I) yielding a SFIFS layer sequence, which allows one to realize much larger values of the product $I_cR_n$, where $I_c$ is the critical current of the junction and $R_n$ - its normal state resistance.\cite{Weides2006, Weides2006_2} Recently, SIFS junctions attracted much attention and have been intensively studied studied both experimentally \cite{Kontos2002, Guichard2003, Born2006, Pepe2006, Weides2006, Weides2006_2, Pfeiffer2008, Bannykh2009, Kemmler2010} and theoretically.\cite{Vasenko2008, Vasenko2011, Buzdin2008, Pugach2009, Volkov2009, Mai2011} For instance, the current-voltage characteristics (CVC) of SIFS Josephson junctions with strong insulating layer were studied in Ref.~\onlinecite{Vasenko2011}. They exhibit interesting nonmonotonic behavior for weak ferromagnetic interlayers, i.e. small enough exchange fields. The reason for this behavior is the shape of the density of states in the F layer. At small exchange fields the decay length of superconducting correlations in ferromagnetic material, $\sim \xi_h$ is large enough, which leads to profound variations of the superconducting density of states in the F layer over energy and results in corresponding CVC behavior. With increase of the exchange field the $\xi_h$ decreases, which suppresses the superconducting correlations in the F layer and makes the SIFS CVC similar to the I-V curve of the FIS junction.

In this paper we study the current-voltage characteristics of SFIFS Josephson junctions with two ferromagnetic interlayers. SFIFS structures were also proposed for various applications in memory elements,\cite{Nevirkovets2018, Nevirkovets2018_2, Shafraniuk2019} single flux quantum (SFQ) circuits,\cite{Shafranjuk2016} and as injectors in superconductor-ferromagnetic transistors (SFT),\cite{Nevirkovets2016, Nevirkovets2014, Nevirkovets2015, Nevirkovets2017} which can be used as amplifiers for memory, digital, and RF applications. In this work we study the current-voltage characteristics of a SFIFS junction, shown in Fig.~\ref{SFmult}. We present quantitative model of the quasiparticle current in SFIFS junctions for different set of parameters characterizing the ferromagnetic interlayers. In case of weak ferromagnetic metals we find the anomalous nonmonotonic shape of the current-voltage characteristics at subgap voltages and compare the results with CVC of SIFS junctions.\cite{Vasenko2011} We ascribe this behavior by DOS energy dependencies in case of small exchange fields in F layers. This shape is smeared if we include finite magnetic scattering rate. The anomalous nonmonotonic shape of the current-voltage characteristics of SFIFS junctions with weak ferromagnetic layers looks similar to the fine structures of quasiparticle currents, recently obtained experimentally on similar systems.\cite{Nevirkovets2014, Nevirkovets2015, Nevirkovets2017, Vavra2017}

The paper organized as follows. In Sec.~\ref{Model} we formulate the theoretical model and basic equations and introduce the self-consistent numerical iterative method for calculating the density of states (DOS) in S/F bilayers. In Sec.~\ref{Results} we present and discuss the results for the density of states in S/F bilayers in case of subgap values of the exchange field [Sec.~\ref{DOS}] and the current-voltage characteristics of SFIFS junctions [Sec.~\ref{CVC}]. Finally we summarize the results in Sec.~\ref{Conclusion}.

%%%%%%%%%%%%%%%%%%%%%%%%%%%%%%%%%%%%%%%%%%%%%%%%%%%%%%%%%%%%%%%%%%%%%%%%%%%%
%%%%%%%%%%%%%%%%%%%%%%%%%%%%%%%%%%%%%%%%%%%%%%%%%%%%%%%%%%%%%%%%%%%%%%%%%%%%

\section{Model}\label{Model}

%%%%%%%%%%%%%%%%%%%%%%%%%%%%%%%%%%%%%%%%%%%%%%%%%%%%%%%%%%%%%%%%%%%%%%%%%%%%
%%%%%%%%%%%%%%%%%%%%%%%%%%%%%%%%%%%%%%%%%%%%%%%%%%%%%%%%%%%%%%%%%%%%%%%%%%%%

%
		\begin{figure}[t]
		\centering
		\includegraphics[width=\columnwidth] {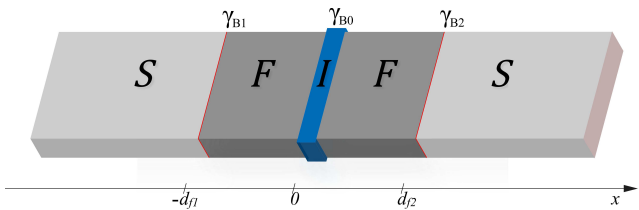}
		\caption
		{(Color online)
			Schematic representation of SFIFS hybrid structure (here S is a superconductor, F is a ferromagnetic metal and I is an insulating barrier). The thicknesses of the ferromagnetic interlayers are $d_{f1}$ and $d_{f2}$, correspondingly. The transparency of the left S/F interface is characterized by $\gamma_{B1}$ parameter, while the transparency of the right F/S interface is characterized by $\gamma_{B1}$ parameter. Both parameters $\gamma_{B1}, \gamma_{B1} \ll 1$, which corresponds to transparent metallic interfaces. The insulating barrier between the left and right interfaces (I) is described by $\gamma_{B0} \gg 1$.
		}
		\label{SFmult}
	\end{figure}

In this section we present the theoretical model we use in our studies. The geometry of the considered system is depicted in Fig.~\ref{SFmult}. It consists of two superconducting electrodes and couple of ferromagnetic interlayers, with thicknesses $d_{f1}$ and $d_{f2}$, correspondingly. The system contains three interfaces: two S/F (superconductor/ferromagnet) boundaries and one tunnel F-I-F interface. Each of these interfaces is described by the dimensionless parameter $\gamma_{Bj}=R_{Bj}\sigma_n/\xi_n$ ($j=0,1,2$), which is proportional to the resistance $R_{Bj}$ across the interface.\cite{KL, gamma_b} Here $\sigma_n$ is the conductivity of the F layer and $\xi_n=\sqrt{D_f/2\pi T_c}$ is the coherence length, where $T_c$ is the critical temperature of the superconductor S (here and below we assume $\hbar = k_B = 1$). In this paper we consider the diffusive limit, when the elastic scattering length $\ell$ is much smaller than the decay characteristic length of the real part of the pair wave function in the ferromagnet, $\xi_{f1}$ [which we introduce later in \Eqs{xif12}]. We assume that the S/F interfaces are not magnetically active. We also neglect the nonequilibrium effects,\cite{VH, Arutyunov2011, Arutyunov2018} and use the Matsubara Green's functions technique, which has been developed to describe many-body systems in equilibrium at finite temperature.\cite{Belzig1999}

In our model the tunneling barrier is located between two F layers at $x=0$ (Fig.~\ref{SFmult}), whereas other interfaces at $x=-d_{f1}$ and $x=d_{f2}$ are identical and transparent. This case corresponds to $\gamma_{B1}=\gamma_{B2} \ll 1$ and  $\gamma_{B0}\gg 1$. In case of strong enough tunnel barrier ($\gamma_{B0}\gg 1$), two S/F bilayers in the SFIFS junction are decoupled, i.e. the amplitudes of two-electron processes between left and right F layers are negligibly small. Hence, the quasiparticle current through the SFIFS junction, biased by the voltage $eV$, can be calculated by using the Werthamer formula,\cite{Werthammer}
	\begin{equation}  \label{Inis}
	I=\frac{1}{e R}\int_{-\infty}^{\infty}dE \; N_{f1}(E-eV) N_{f2}(E) [f(E-eV)-f(E)],
	\end{equation}
where $N_{f1,2}(E)$ is the density of states (DOS) in the corresponding ferromagnetic layer at $x=0$, $f(E)=[1+e^{E/T}]^{-1}$ is the Fermi-Dirac distribution function, and $R=R_{B0}$ is the resistance across the F-I-F interface. Both densities of states $N_{f1,2}(E)$ are normalized to their values
in the normal state.
	
In order to obtain the densities of states in ferromagnetic layers, $N_{f1,2}(E)$, we use a self-consistent two-step iterative procedure, described below. As far as $\gamma_{B0} \gg 1$, we can neglect the influence of right F layer on the density of states in the left S/F bilayer and vice versa (see Fig.~\ref{SFmult}). Thus we need to obtain the DOS at the outer border of each S/F bilayer. That can be done by solving the Usadel equations in S/F bilayer system.\cite{Usadel}

In the following, we use the $\theta$-parameterizations of normal ($G=\cos \theta$) and anomalous ($F=\sin \theta$) Green's functions and write the Usadel equations in F layers in the form,\cite{Usadel, Gusakova2006}
\begin{align}
\frac{D_f}{2} \frac{\partial ^{2}\theta _{f\uparrow (\downarrow)}}{\partial x^{2}} = &\left( \omega \pm ih + \frac{1}{\tau_z}\cos\theta_{f\uparrow (\downarrow )}\right) \sin\theta_{f\uparrow (\downarrow)} \nonumber
\\
&+ \frac{1}{\tau_x} \sin(\theta_{f\uparrow} + \theta_{f\downarrow}) \pm \frac{1}{\tau_{so}} \sin(\theta_{f\uparrow} - \theta_{f\downarrow}),
\end{align}
where the positive and negative signs correspond to the spin-up (``$\uparrow$'') and spin-down (``$\downarrow$'') states, respectively. In terms of the electron fermionic operators $\psi_{\uparrow (\downarrow)}$ the spin-up state corresponds to the anomalous Green's function $F_\uparrow \sim \langle \psi_\uparrow \psi_\downarrow \rangle$, while spin-down state corresponds to $F_\downarrow \sim \langle \psi_\downarrow \psi_\uparrow \rangle$. The
$\omega = 2 \pi T (n + \frac{1}{2})$ are the Matsubara frequencies, where $n=0,\pm 1, \pm2, \ldots$, and $h$ is the exchange field in the ferromagnet. The scattering times are labeled here as $\tau_z$, $\tau_x$, and $\tau_{so}$, where $\tau_{z(x)}$ corresponds to the magnetic scattering parallel (perpendicular) to the quantization axis, and $\tau_{so}$ is the spin-orbit scattering time.\cite{Abrikosov1960, Faure2006, Bergeret2007, Ivanov2009}

Assuming strong uniaxial anisotropy in ferromagnetic materials, in which case there is no coupling between spin-up and spin-down electron populations, we neglect $\tau_x$ ($\tau^{-1}_x \sim 0$). Moreover we also assume the ferromagnets with weak spin-orbit coupling and thus neglect spin-orbit scattering time $\tau_{so}$. After taking into account all the assumptions mentioned above the Usadel equations in the ferromagnetic layers for different spin states can be written as
	\begin{equation}\label{Usadel}
	\frac{D_{f}}{2} \frac{\partial ^{2}\theta _{f\uparrow (\downarrow
			)}}{\partial x^{2}} = \left( \omega \pm ih + \frac{\cos
		\theta_{f\uparrow (\downarrow )}}{\tau _{m}}\right) \sin
	\theta_{f\uparrow (\downarrow )},
	\end{equation}
where $\tau_m \equiv \tau_z$ is the magnetic scattering time. In the superconducting layer S the Usadel equation read\cite{Usadel}
	\begin{equation}\label{Usadel_S}
	\frac{D_s}{2} \frac{\partial^2 \theta_s}{\partial x^2} = \omega
	\sin \theta_s - \Delta(x) \cos \theta_s.
	\end{equation}
Here $D_s$ is the diffusion coefficient in the S layer and $\Delta(x)$ is the pair potential in the superconductor. We note that $\Delta(x)$ vanishes in the F layer.

\Eqs{Usadel} and \eqref{Usadel_S} must be supplemented with corresponding boundary conditions. At the S/F interfaces we apply the Kupriyanov-Lukichev boundary conditions. For example, at the left S/F interface they are written as,\cite{KL}
\begin{subequations}
	\label{KL}
	\begin{align}
	\xi_n\gamma\left( \frac{\partial \theta_f}{\partial x} \right)_{-d_{f1}} &=
	\xi_s \left( \frac{\partial \theta_s}{\partial x} \right)_{- d_{f1}},
	\label{KL1} \\
	\xi_n \gamma_{B1} \left( \frac{\partial \theta_f}{\partial x}
	\right)_{ - d_{f1}} &= \sin\left( \theta_s - \theta_f \right)_{ - d_{f1}}.
	\label{KL_DOS}
	\end{align}
\end{subequations}
Similar equations can be written at the right S/F interface at $x=d_{f2}$. Here $\gamma = \xi_s\sigma_n/\xi_n\sigma_s$, where $\sigma_s$ is the conductivity of the S layer and $\xi_s = \sqrt{D_s/2\pi T_c}$ is the superconducting coherence length. The parameter $\gamma$ defines the strength of the inverse proximity effect, i.e. suppression of superconductivity in the adjacent S layer by the ferromagnetic layer F. We consider the parameter $\gamma$ to be relatively small $\gamma \ll 1$, which corresponds to rather weak suppression.

To calculate the density of states in the S/F bilayer we should set the boundary conditions at the  outer  boundary of the
ferromagnet ($x=0$),
\begin{equation}\label{leftBK}
\left(\frac{\partial \theta_f}{\partial x} \right)_{0} = 0.
\end{equation}
To complete the boundary problem we also set a boundary condition at $x=\pm\infty$,
\begin{equation}\label{gran1}
\theta_s(\pm\infty) = \arctan\frac{\Delta}{\omega},
\end{equation}
where the Green's functions acquire the well-known bulk BCS form. We notice that the density
of states at $x=\pm\infty$ is given by standard BCS equation,
\begin{equation}
N_s(E) = \re \left [ \cos \theta_s (i\omega \rightarrow E + i0) \right ] = \frac{|E| \Theta(|E|-\Delta)}{\sqrt{E^2 - \Delta^2}},
\end{equation}
where $\Theta(x)$ is the Heaviside step function.

Finally the self-consistency equation for the superconducting order parameter takes the form,
	\begin{equation}
	\Delta (x)\ln \frac{T_c}{T} = \pi T \sum\limits_{\omega > 0} \left( \frac{2\Delta
		(x)}{\omega}-\sin \theta _{s \uparrow} - \sin \theta _{s \downarrow} \right)
	\label{Delta}.
	\end{equation}
The equations \eqref{Usadel}-\eqref{gran1} and \Eq{Delta} represent a closed set of equations that should be solved self-consistently.

The density of states $N_{f1,2}(E)$ normalized to the DOS in the normal state, can be written as
 \begin{equation}
 N_{fj}(E) = \left[ N_{fj \uparrow}(E) + N_{fj \downarrow}(E)\right]/2, \quad j=1,2,
 \label{DOS_full}
 \end{equation}
 where $N_{fj \uparrow(\downarrow)}(E)$ are the spin resolved densities of states written in terms of the spectral angle $\theta$,
 \begin{equation}
 N_{fj \uparrow(\downarrow)}(E) =
 \re\left[\cos\theta_{fj \uparrow(\downarrow)}(%
 i\omega \rightarrow E + i0)\right], \quad j=1,2.\label{DOS_spin}
 \end{equation}

To obtain $N_{f1,2}$, we use a self-consistent two-step iterative procedure.\cite{Gusakova2006, Golubov2002, Golubov1988, Golubov1995}
In the first step we calculate the pair
potential coordinate dependence $\Delta(x)$ using the self-consistency equation in the S layer, Eq.~(\ref{Delta}). Then, by proceeding to the analytical continuation in Eqs.~(\ref{Usadel}),~\eqref{Usadel_S} over the quasiparticle energy $i\omega \rightarrow
E + i0$ and using the $\Delta(x)$ dependence obtained in the previous step, we find the Green's functions by repeating the iterations until convergency is reached.

The characteristic lengths of the decay and oscillations of the real part of the pair wave function in the ferromagnetic layer at the Fermi energy, $\xi_{f1,2}$, are given in our model by,\cite{Vasenko2011}
\begin{subequations}\label{xif12}
\begin{align}
\frac{1}{\xi_{f1}} &= \frac{1}{D_f} \sqrt{\sqrt{h^2 + \frac{1}{\tau_m^2}} + \frac{1}{\tau_m}},
\\
\frac{1}{\xi_{f2}} &= \frac{1}{D_f} \sqrt{\sqrt{h^2 + \frac{1}{\tau_m^2}} - \frac{1}{\tau_m}}.
\end{align}
\end{subequations}
We see from these equations that with increase of the magnetic scattering rate $\alpha_m = 1/\tau_m \Delta$
the length of decay $\xi_{f1}$ decreases, while the
length of oscillations $\xi_{f2}$ increases. In the absence of magnetic scattering $\xi_{f1} = \xi_{f2} = \xi_h = \sqrt{D_f/h}$.

	\begin{figure}[t]
		\centering
		\includegraphics[width=\columnwidth] {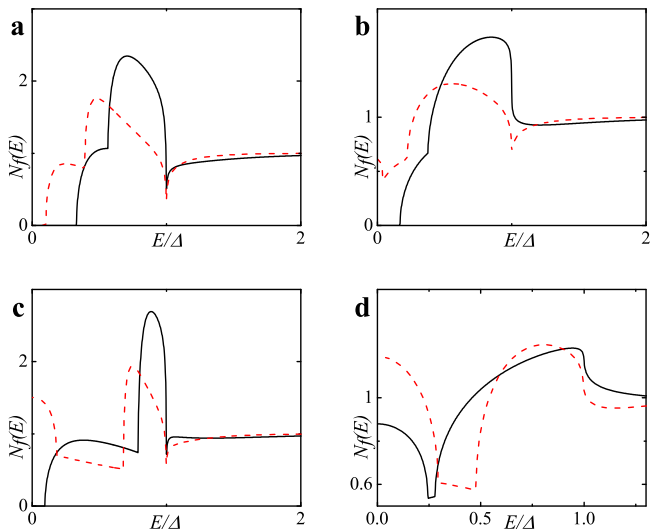}
		\caption
		{(Color online)
			DOS $N_f(E)$ on the free boundary of the F layer in the FS bilayer obtained numerically for two cases: (a) in the absence of magnetic scattering, $\alpha_m = 1/\tau_m \Delta =0$ (plots a and c) and in case of finite magnetic scattering - plot b ($\alpha_m=0.1$) and plot d ($\alpha_m=0.5$). Parameters of the FS interface are $\gamma=\gamma_B=0.01$, and $T=0.1T_c$. Plots a-b: $h = 0.1\Delta$; plots c-d: $h = 0.3\Delta$. Black solid line corresponds to $d_f=2\xi_n$, while red dashed line to $d_f=3\xi_n$.
		}
		\label{DOS_1}
	\end{figure}
	\begin{figure}[t]
	\centering
	\includegraphics[width=\columnwidth] {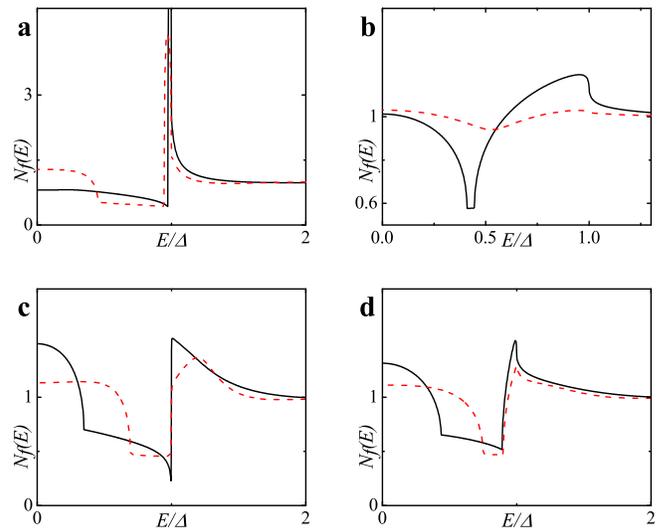}
	\caption
	{(Color online)
		DOS $N_f(E)$ on the free boundary of the F layer in the FS bilayer obtained numerically in the absence of magnetic scattering, $\alpha_m = 1/\tau_m \Delta = 0$ (plots a and c) and in case of finite magnetic scattering - plot d ($\alpha_m=0.1$) and plot b ($\alpha_m=0.5$). Plots a-b: $h = 0.5\Delta$; plots c-d: $h = 0.7\Delta$. Black solid line corresponds to $d_f=2\xi_n$, while red dashed line to $d_f=3\xi_n$.
	}
	\label{DOS_2}
\end{figure}
 \begin{figure}[h!]
 	\centering
 	\includegraphics[width=\columnwidth] {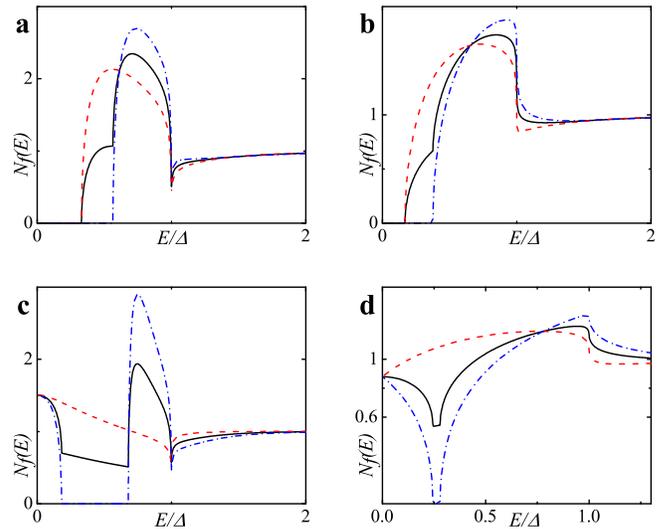}
 	\caption
 	{(Color online)
 		Spin resolved DOS $N_{f\uparrow(\downarrow)}$ on the free boundary of the F layer in the FS bilayer calculated numerically in the absence of magnetic scattering, $\alpha_m=0$  (plots a and c) and in case of finite magnetic scattering - plot b ($\alpha_m=0.1$) and plot d ($\alpha_m=0.5$). Plots a-b: $h = 0.5\Delta$, $d_f=2\xi_n$; plots c-d: $h = 0.3\Delta$, $d_f=3\xi_n$ (c) and $d_f=2\xi_n$(d). Black solid line corresponds to $N_{f}(E)$, red dashed line to $N_{f\uparrow}(E)$ and blue dash-dotted line to $N_{f\downarrow}(E)$.
 	}
 	\label{DOS_resolved}
 \end{figure}
%

%%%%%%%%%%%%%%%%%%%%%%%%%%%%%%%%%%%%%%%%%%%%%%%%%%%%%%%%%%%%%%%%%%%%%%%%%%%%
%%%%%%%%%%%%%%%%%%%%%%%%%%%%%%%%%%%%%%%%%%%%%%%%%%%%%%%%%%%%%%%%%%%%%%%%%%%%

\section{Results and discussion}\label{Results}

%%%%%%%%%%%%%%%%%%%%%%%%%%%%%%%%%%%%%%%%%%%%%%%%%%%%%%%%%%%%%%%%%%%%%%%%%%%%
%%%%%%%%%%%%%%%%%%%%%%%%%%%%%%%%%%%%%%%%%%%%%%%%%%%%%%%%%%%%%%%%%%%%%%%%%%%%

In this section we present the results of the DOS energy dependencies in SF bilayers at free boundary of the F layer for $h \lesssim \Delta$ [Sec.~\ref{DOS}]. The densities of states for $h \gtrsim \Delta$ were thoroughly discussed in Ref.~\onlinecite{Vasenko2011}. Then we calculate corresponding CVC of the SFIFS junction using the Werthamer formula, \Eq{Inis}. In case of $h \lesssim \Delta$ we obtain interesting nonmonotonic behavior of the quasiparticle current, presented in Sec.~\ref{CVC}. At large exchange fields the decay length $\xi_{f2}$ of the real part of the pair wave function in the F layer became small [see \Eqs{xif12}] and the amplitude of DOS variations tends to zero. In this case the CVC of SFIFS junction tends to Ohm's law for $h \gg \Delta$. The ferromagnetic materials with small exchange fields can be fabricated as discussed in Ref.~\onlinecite{smallfield}. We also note that the DOS at the end of an SF bilayer in case of the domain wall in the ferromagnetic layer was studied in Ref.~\onlinecite{Bobkova2019}.

%%%%%%%%%%%%%%%%%%%%%%%%%%%%%%%%%%%%%%%%%%%%%%%%%%%%%%%%%%%%%%%%%%%%%%%%%%%%

\subsection{Density of states in SF bilayers for $h \lesssim \Delta$}\label{DOS}

%%%%%%%%%%%%%%%%%%%%%%%%%%%%%%%%%%%%%%%%%%%%%%%%%%%%%%%%%%%%%%%%%%%%%%%%%%%%

Figures \ref{DOS_1} and \ref{DOS_2} show the DOS energy dependencies for different $h \lesssim \Delta$ and for relatively thick F layers. In our calculations we fix the temperature at $T=0.1 T_c$, where $T_c$ is the critical temperature of the superconductor S. In Fig.~\ref{DOS_1} the characteristic ``finger-like'' shape of DOS is observed along with a minigap for $d_f=2\xi_n$ [Fig.~\ref{DOS_1} (a) and (c)]. At larger $d_f$ as and/or at larger $h$ the minigap closes [Fig.~\ref{DOS_1} (c) and Fig.~\ref{DOS_2} (a, c)]. In the absence of magnetic scattering ($\alpha_m = 1/\tau_m \Delta = 0$) we can roughly estimate the critical value $h_c$ of the exchange field at which the minigap closes as\cite{Vasenko2011}
\begin{equation}
h_c \sim E_{Th}, \quad E_{Th} = D_f/d_f^2,
\end{equation}
where $E_{Th}$ is the Thouless energy and $d_f$ is the thickness of the F layer in the SF bilayer [$d_{f1}$ or $d_{f2}$ for the left or right SF bilayer in Fig.~\ref{SFmult}]. Since we consider subgap values of $h$, the minigap closes at rather large $d_f$ in the absence of magnetic scattering.

After the minigap closes the DOS at the Fermi energy $N_f(0)$ rapidly increases to values larger than unity with further increase of $d_f$ and then it oscillates around unity while its absolute value exponentially approaches unity.\cite{Vasenko2011} This is the well-known damped oscillatory behavior with the lengthes of decay and oscillations given by \Eqs{xif12}, correspondingly. Figures \ref{DOS_1} (b, d) and \ref{DOS_2} (b, d) show that stronger magnetic scattering leads to the minigap closing at smaller $d_f$.
With the increase of $\alpha_m = 1/\tau_m \Delta$ the period of oscillations increases [$\xi_{f2}$ in \Eqs{xif12} increases].
At the same time the DOS variation amplitude became smaller and DOS features smear, since for larger $\alpha_m$ the dumped exponential decay of oscillations occurs faster [$\xi_{f1}$ in \Eqs{xif12} decreases].

Finally, we present plots for spin-resolved densities of states given by \Eqs{DOS_spin} in Fig.~\ref{DOS_resolved} for both zero and finite magnetic scattering.

\begin{figure}[t]
\centering
\includegraphics[width=\columnwidth] {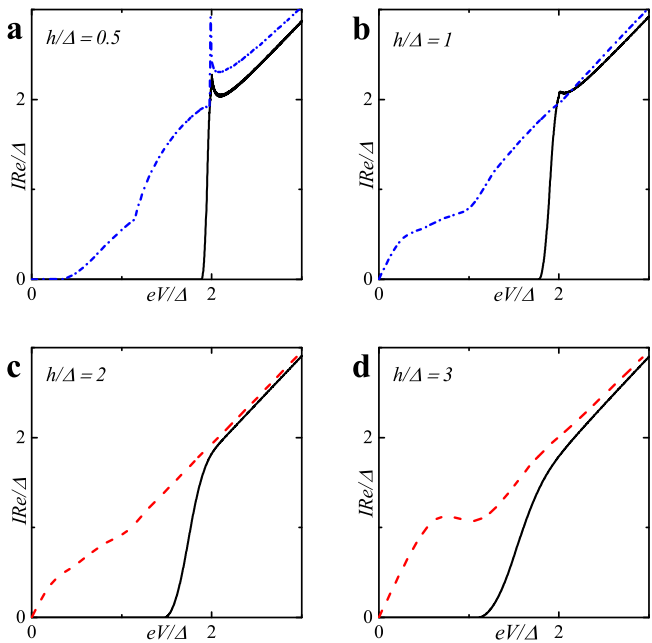}
\caption
{(Color online)
Current-voltage characteristics of the symmetric ($d_{f1}=d_{f2}=d_f$) SFIFS junction in the absence of magnetic scattering for different values of exchange field $h$. The temperature $T = 0.1T_c$. In each graph the curves were calculated for different values of F layer thickness $d_f$,  $d_f=0.5\xi_n$ (black solid line), $d_f=1.0\xi_n$ (red dashed line), $d_f=1.5\xi_n$ (blue dash-dotted line).
}
\label{SFIFS_1}
\end{figure}
\begin{figure}[t]
	\centering
	\includegraphics[width=\columnwidth] {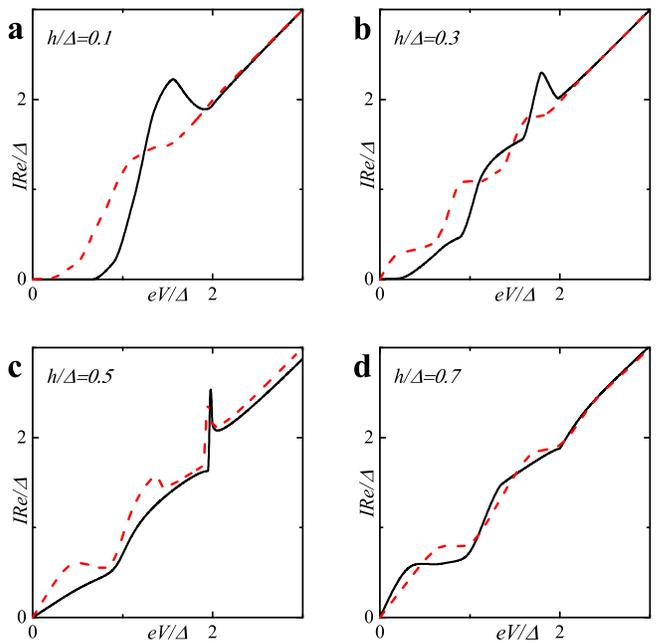}
	\caption
	{(Color online)
Current-voltage characteristics of a symmetric SFIFS junction for different values of subgap exchange field $h$ in the absence of magnetic scattering. The temperature $T = 0.1T_c$. In each graph the curves were calculated for different values of F layer thickness $d_f$, $d_f=2\xi_n$ (black solid line) and $d_f=3\xi_n$ (red dashed line).
	}
	\label{SFIFS_3}
\end{figure}
\begin{figure}[t]
	\centering
	\includegraphics[width=\columnwidth] {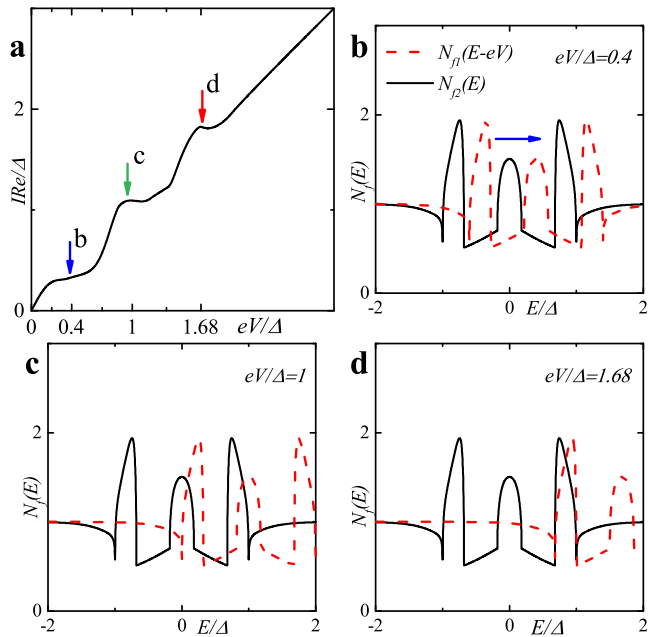}
	\caption
	{(Color online)
		 The CVC taken from Fig.~\ref{SFIFS_3} (b), red dashed line, and visual explanation of the characteristic behavior of the quasiparticle current (a). Plots (b)-(d) show the DOS $N_{f}(E-eV)$ and $N_{f}(E)$ at particular value of $eV$ revealing the origin of the current features in plot (a).
	}
	\label{SFIFS_expl}
\end{figure}
\begin{figure}[t]
	\centering
	\includegraphics[width=\columnwidth] {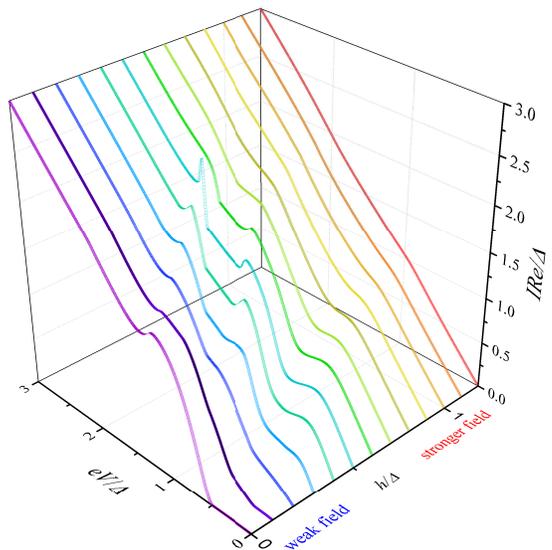}
	\caption
	{(Color online) Current-voltage characteristics of a symmetric SFIFS junction in the absence of magnetic scattering for $d_f=3\xi_n$. The temperature $T = 0.1T_c$. The curves correspond to different values of $h$, from $h=0\Delta$ to $h=1.2\Delta$ with increment equal to $0.1\Delta$. The exchange field $h=0$ corresponds to the case of a SNINS junction.\cite{Golubov1988}
	}
	\label{SFIFS_3d}
\end{figure}
\begin{figure}[t]
	\centering
	\includegraphics[width=\columnwidth] {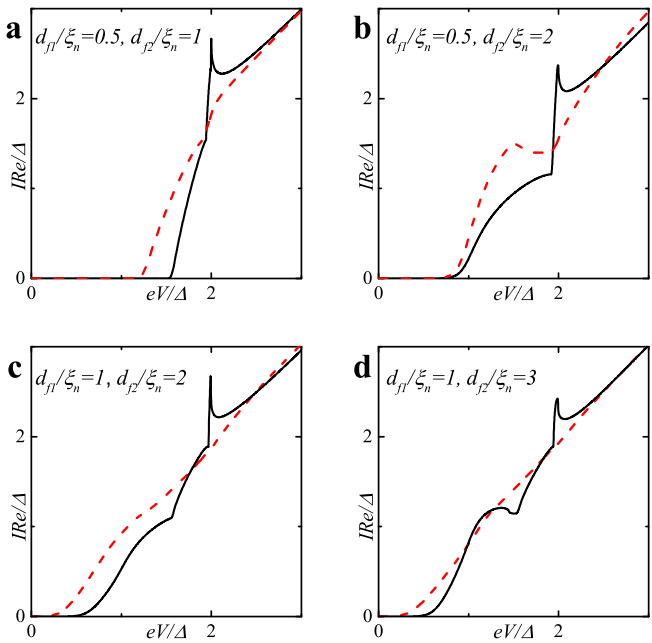}
	\caption
	{(Color online)
Current-voltage characteristics of an asymmetric ($d_{f1}\neq d_{f2}$) SFIFS junction for different values of F layer thicknesses $d_{f1}$ and $d_{f2}$ (indicated in the plot) in the absence of magnetic scattering. The temperature $T = 0.1T_c$, $h=0.5\Delta$ (black solid line) and $h=1.0\Delta$ (red dashed line).
	}
	\label{SFIFS_2}
\end{figure}
\begin{figure}[t]
	\centering
	\includegraphics[width=\columnwidth] {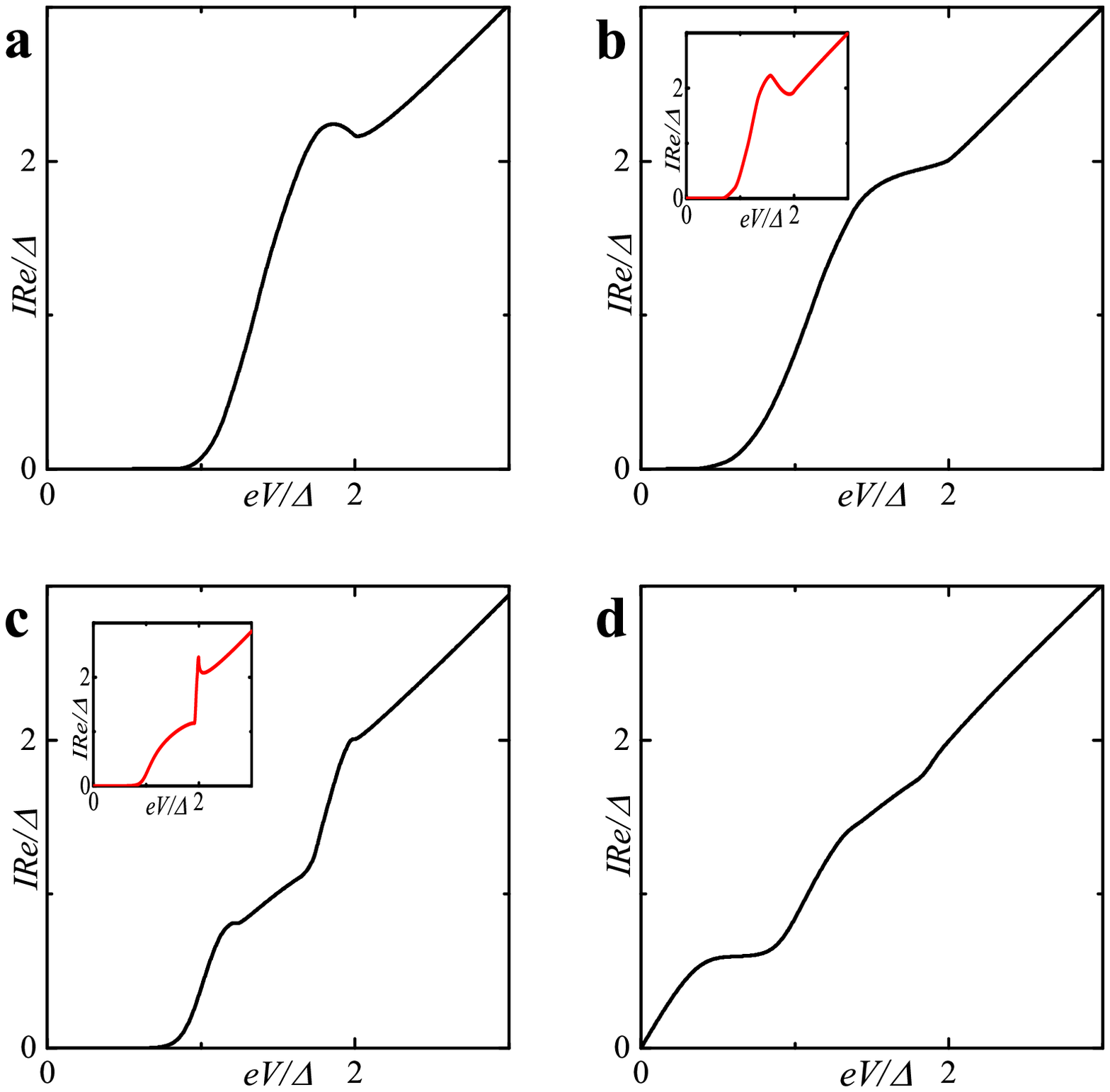}
	\caption
	{(Color online)
		Current-voltage characteristics of a SFIFS junction in the presence of magnetic scattering ($\alpha_m=0.1$). The temperature $T = 0.1T_c$. In the plot (a) black solid line corresponds to $d_{f1}=1\xi_n, d_{f2}=2\xi_n$, in the plots (b) and (d) to $d_{f1}=d_{f2}=2\xi_n$ and finally in the plot (c) black line corresponds to $d_{f1}=0.5\xi_n, d_{f2}=2\xi_n$. Plots (a)-(b): $h=0.1\Delta$; plots (c) and (d): $h=0.5\Delta$ and $h=0.7\Delta$, respectively. The insets show the CVC in case of zero magnetic scattering.
	}
	\label{SFIFS_scat}
\end{figure}
%

%%%%%%%%%%%%%%%%%%%%%%%%%%%%%%%%%%%%%%%%%%%%%%%%%%%%%%%%%%%%%%%%%%%%%%%%%%%%

\subsection{Current-voltage characteristics of SFIFS junctions}\label{CVC}

%%%%%%%%%%%%%%%%%%%%%%%%%%%%%%%%%%%%%%%%%%%%%%%%%%%%%%%%%%%%%%%%%%%%%%%%%%%%

Using the densities of states $N_{f1,2}(E)$ obtained in Sec.~\ref{DOS}, we calculate a set of quasiparticle current curves using \Eq{Inis} for various values of parameters describing properties of ferromagnetic material, which include F layer thicknesses $d_{f1}$ and $d_{f2}$, exchange field $h$, and magnetic scattering rate $\alpha_m$. In our calculations we fix the temperature at $T=0,1T_c$, where $T_c$ is the critical temperature of the superconducting lead.

Fig.~\ref{SFIFS_1} demonstrates the CVC of a symmetric SFIFS junction, where $d_{f1}=d_{f2}=d_f$ in the absence of magnetic scattering. For thin enough ferromagnetic interlayers, $d_f/\xi_n = 0.5$, and small enough value of the exchange field, $h=0.5\Delta$, we observe the CVC which resemble the I-V characteristic of a SNINS Josephson junction with a characteristic peak at $eV \approx 2\Delta$ [see Fig.~\ref{SFIFS_1} (a), solid black line].\cite{Golubov1988} With increase of the exchange field $h$ this peak is smeared [see Fig.~\ref{SFIFS_1} (b), (c) and (d), solid black line]. Increasing the $d_f$ and/or $h$ produce a set of I-V curves, among which the red dashed line in Fig.~\ref{SFIFS_1} (d) is the most interesting, since it performs a nonmonotonic behavior. The reason of a typical nonmonotonic behavior will be explained later.

Fig.~\ref{SFIFS_3} shows the current-voltage characteristics of SFIFS junctions at subgap values of the exchange field. We observe a nonmonotonic behavior for thick enough ferromagnetic layers at $h \lesssim \Delta$. Let us consider the CVC in Fig.~\ref{SFIFS_3} (b), red dashed line. We can explain its behavior as well as any other nonmonotonic CVC behavior as the signature of the DOS energy dependence. The anomalous nonmonotonic I(V) dependence arises from the shape features of the densities of states, see Fig.~\ref{SFIFS_expl}. In symmetric SFIFS junctions, $N_{f1}(E) = N_{f2}(E) \equiv N_f(E)$ in \Eq{Inis}, which
can be well approximated by taking $T=0$ for small temperatures $T \ll T_c$. In this case the Fermi-Dirac distribution function $f(E)$ can be represented as the Heaviside step function $\Theta(-E)$ [and $f(E-eV)$ as $\Theta(eV-E)$]. As a result, the limits of integration in (\ref{Inis}) shrink to the interval $[0,eV]$. Hence, the current through the junction can be written as,
	\begin{equation}  \label{Ifif2}
	I=\frac{1}{e R}\int_{0}^{eV}dE N_{f}(E-eV) N_{f}(E).
	\end{equation}

Using this expression, the origin of nonmonotonic behavior of the CVC can be explained. At $eV=0$ the upper limit of the integral in \Eq{Ifif2} is zero and the current is zero. With the increase of the voltage, the current first increases linearly due to broader region of integration as in Ohm's law. The first feature which is shown on Fig.~\ref{SFIFS_expl} (a) is a significant change in the slope of the current. Fig.~\ref{SFIFS_expl} (b) shows relative positions of the densities of states $N_{f}(E-eV)$ and $N_{f}(E)$ in this case, where almost no peak overlap can be seen, resulting in relatively small value of the integral in \Eq{Ifif2}. As we proceed to larger values of $eV$, we reach the first local maximum of the CVC which corresponds to maximum overlap of the densities of states $N_{f}(E-eV)$ and $N_{f}(E)$ at $eV/\Delta \approx 1$ [see Fig.~\ref{SFIFS_expl} (c)]. The second maximum of the quasiparticle current occurs at $eV/\Delta\approx1.68$ that corresponds to perfect DOS peak overlap at $E/\Delta\approx 1$ [Fig.~\ref{SFIFS_expl} (d)]. For large enough values of voltage $eV$, a product of the DOS $N_{f}(E-eV) N_{f}(E) \approx 1$ and its integration does not produce any features. Thus, the CVC eventually coincides with Ohm's law in this case. In fact any shape of a SFIFS I-V curve can be explained and understood in this way. We note that in this paper we present the densities of states in SF bilayers only for subgap values of the exchange field. For $h \gtrsim \Delta$ the DOS energy dependencies in SF bilayers can be found, for example, in Ref.~\onlinecite{Vasenko2011}.

Based on the properties of the density of states in FS bilayers we can see that even the tiny exchange field $h$ can modify the current dramatically introducing anomalous nonmonotonic behavior in case of thick enough F layers [see Figs.~\ref{SFIFS_1}, \ref{SFIFS_3}]. It is important then to understand how the CVC of a SFIFS junction transforms as the exchange field $h$ increases. In Fig.~\ref{SFIFS_3d} we demonstrate the plot of current-voltage characteristics calculated for a wide range of exchange field values $h$ in the absence of magnetic scattering. From this plot it can be clearly seen that while for relatively small (subgap) values of the exchange field many interesting features appear in the structure of the current, at larger values of $h$ these features are smeared and CVC tends to the Ohm's law. Figure \ref{SFIFS_2} shows the current-voltage characteristics in case of an asymmetric SFIFS junction, i.e. when $d_{f1} \ne d_{f2}$ in case of zero magnetic scattering.

In this section we also present the current-voltage characteristics of a SFIFS junction calculated in the presence of magnetic scattering for different values of the subgap exchange field $h$. Fig.~\ref{SFIFS_scat} illustrates the CVC in case of finite magnetic scattering rate $\alpha_m=0.1$. We consider both symmetric and asymmetric SFIFS junctions. The insets show the CVC in case of zero magnetic scattering. For tiny $h$ nonzero magnetic scattering leads to smearing of characteristic features of the current as shown in Fig.~\ref{SFIFS_scat}. At larger subgap values of the exchange field $h$ we see a ``triple kink'' structure, see Fig.~\ref{SFIFS_scat} (c). For large enough values of $\alpha_m$ the nonmonotonic behavior of the quasiparticle current will be smeared and the current tends to the Ohm's law. This is due to the fact that increasing $\alpha_m$ the length of the superconducting correlations decay in the ferromagnetic layers decreases, see \Eqs{xif12}, and the supression of superconducting correlations in the F layers occurs faster.

We can compare these results with the I-V characteristics of SIFS Josephson junctions.\cite{Vasenko2011} In this case at zero magnetic scattering we may also observe the nonmonotonic behavior, but with only one peak [see Ref.~\onlinecite{Vasenko2011}, Fig. 6 (c)]. In case of finite magnetic scattering the CVC has a ``double kink'' structure [see Ref.~\onlinecite{Vasenko2011}, Fig. 7 (a, c)]. In SFIFS junctions the overlap of subgap DOS structures $N_{f1}(E-eV) N_{f2}(E)$ in the integrand of the current equation, \Eq{Ifif2}, produce more complex behavior of the I-V characteristics.

We also notice that in recent experiments on SFIFS junctions as injectors of superconductor-ferromagnetic transistors (SFT) some fine structures of the subgap quasiparticle current was observed,\cite{Nevirkovets2014, Nevirkovets2015, Nevirkovets2017, Vavra2017} which looks similar to our theoretical results.

%%%%%%%%%%%%%%%%%%%%%%%%%%%%%%%%%%%%%%%%%%%%%%%%%%%%%%%%%%%%%%%%%%%%%%%%%%%%
%%%%%%%%%%%%%%%%%%%%%%%%%%%%%%%%%%%%%%%%%%%%%%%%%%%%%%%%%%%%%%%%%%%%%%%%%%%%

\section{Conclusion}\label{Conclusion}

%%%%%%%%%%%%%%%%%%%%%%%%%%%%%%%%%%%%%%%%%%%%%%%%%%%%%%%%%%%%%%%%%%%%%%%%%%%%
%%%%%%%%%%%%%%%%%%%%%%%%%%%%%%%%%%%%%%%%%%%%%%%%%%%%%%%%%%%%%%%%%%%%%%%%%%%%

In this work we have presented the results of CVC calculations of a SFIFS junction for different set of parameters including the thicknesses of ferromagnetic layers $d_{f1}, d_{f2}$, the exchange field, and the magnetic scattering time $\alpha_m = 1/\tau_m \Delta$. We considered the case of a strong insulating barrier such that the left SF and the right FS bilayers are decoupled. In order to obtain the current-voltage characteristics we first calculated the densities of states (DOS) on the free boundary of the F layer in each SF bilayer utilizing the iterative self-consistent approach. Using the numerically calculated DOS we have derived the quasiparticle current of a SFIFS junction in the case of symmetric ($d_{f1}=d_{f2}$) and asymmetric ($d_{f1} \ne d_{f2}$) structures. We have paid much attention to the case of SFIFS junction with weak ferromagnetic interlayers with exchange fields $h \lesssim \Delta$. It was demonstrated that the CVC possess interesting and unusual features in this case, which can be ascribed by typical DOS behavior. We have provided simple physical explanation of the CVC with such anomalous behavior. We have also illustrated how the CVC shape evolves as one increases the exchange field $h$ introducing. It should be emphasized that taking into account finite magnetic scattering leads to the smearing of characteristic features and in particular cases leads to a ``triple kink'' shape of the current.

%%%%%%%%%%%%%%%%%%%%%%%%%%%%%%%%%%%%%%%%%%%%%%%%%%%%%%%%%%%%%%%%%%%%%%%%%%%%
%%%%%%%%%%%%%%%%%%%%%%%%%%%%%%%%%%%%%%%%%%%%%%%%%%%%%%%%%%%%%%%%%%%%%%%%%%%%

\acknowledgements

%%%%%%%%%%%%%%%%%%%%%%%%%%%%%%%%%%%%%%%%%%%%%%%%%%%%%%%%%%%%%%%%%%%%%%%%%%%%
%%%%%%%%%%%%%%%%%%%%%%%%%%%%%%%%%%%%%%%%%%%%%%%%%%%%%%%%%%%%%%%%%%%%%%%%%%%%

The authors thank D. Beckmann for useful discussions. S.K. acknowledge the hospitality of the Quantum nanoelectronics laboratory of Moscow Institute of Electronics and Mathematics in National Research University Higher School of Economics during his stay in Moscow.

%%%%%%%%%%%%%%%%%%%%%%%%%%%%%%%%%%%%%%%%%%%%%%%%%%%%%%%%%%%%%%%%%%%%%%%%%%%%
%%%%%%%%%%%%%%%%%%%%%%%%%%%%%%%%%%%%%%%%%%%%%%%%%%%%%%%%%%%%%%%%%%%%%%%%%%%%

\end{document}